\documentstyle[12pt]{article}
\newlength{\extraspace}
\newlength{\extraspaces}
\newcommand{\be}{\begin{equation}
\addtolength{\abovedisplayskip}{\extraspaces}
\addtolength{\belowdisplayskip}{\extraspaces}
\addtolength{\abovedisplayshortskip}{\extraspace}
\addtolength{\belowdisplayshortskip}{\extraspace}}
\newcommand{\ee}{\end{equation}}
\newcommand{\ba}{\begin{eqnarray}
\addtolength{\abovedisplayskip}{\extraspaces}
\addtolength{\belowdisplayskip}{\extraspaces}
\addtolength{\abovedisplayshortskip}{\extraspace}
\addtolength{\belowdisplayshortskip}{\extraspace}}
\newcommand{\ea}{\end{eqnarray}}
\newcommand{\nonu}{\nonumber \\[.5mm]}
\newcommand{\A}{&\!\!\!}
\begin{document}
\thispagestyle{empty}
\begin{flushright}
SIT-LP-03/10 \\
{\tt hep-th/0310097} \\
October, 2003
\end{flushright}
\vspace{7mm}
\begin{center}
{\large \bf Linearizing $N = 1$ nonlinear supersymmetry 
with higher derivative terms of a Nambu-Goldstone fermion 
} \\[20mm]
{\sc Kazunari Shima}
\footnote{
\tt e-mail: shima@sit.ac.jp} \ 
and \ 
{\sc Motomu Tsuda}
\footnote{
\tt e-mail: tsuda@sit.ac.jp} 
\\[5mm]
{\it Laboratory of Physics, 
Saitama Institute of Technology \\
Okabe-machi, Saitama 369-0293, Japan} \\[20mm]
\begin{abstract}
We investigate for $N = 1$ supersymmetry (SUSY) 
the relation between a scalar supermultiplet of linear SUSY 
and a nonlinear (NL) SUSY model including apparently pathological 
higher derivaive terms of a Nambu-Goldstone (N-G) fermion 
besides the Volkov-Akulov (V-A) action. 
SUSY invariant relations with higher derivative terms 
of the N-G fermion, which connect the linear and NL SUSY models, 
are constructed at leading orders by heuristic arguments. 
We discuss a higher derivative action 
of the N-G fermion in the NL SUSY model, 
which apparently includes a (Weyl) ghost field. 
By using this relation, 
we also explicitly prove an equivalence between the standard 
NL SUSY V-A model and our NL SUSY model 
with the pathological higher derivatives 
as an example with respect to the universality 
of NL SUSY actions with the N-G fermion. 

/Keywords: supersymmetry, Nambu-Goldstone fermion 
\end{abstract}
\end{center}

\newpage

There are two different realizations of supersymmetry (SUSY), 
the linear realization \cite{WZ1}, and the nonlinear realization \cite{VA} 
which characterize Nambu-Goldstone (N-G) fermions \cite{SS} 
indicating the spontaneously SUSY breaking (SSB) \cite{FI,O}. 
The relation between linear and NL SUSY models was investigated 
by many authors \cite{IK}-\cite{STT2} as an analogy with internal symmetries. 
Indeed, for $N = 1$ SUSY it is well-known \cite{IK}-\cite{UZ} 
that the Volkov-Akulov (V-A) model \cite{VA} of NL SUSY 
is algebraicly equivalent to a scalar supermultiplet 
of linear SUSY \cite{WZ1}. 
The relationship between the V-A model and a (axial vector) gauge supermultiplet 
with the Fayet-Iliopoulos (F-I) $D$ term indicating SSB was also studied 
in \cite{IK,STT1}. For $N = 2$ SUSY we have proved 
by heuristic arguments \cite{STT2} that the V-A model is equivalent 
to a (vector) gauge supermultiplet 
with general F-I $D$ terms, which has a SU(2) ($\times$ U(1)) symmetry. 

On the other hand, a (NL SUSY invariant) higher derivative action 
of the N-G fermion besides the V-A action was explicitly shown 
in the superspace formalism (the construction of V-A superfield) \cite{SW}, 
while in the model towards the SUSY composite unified model 
which is called the superon-graviton model (SGM) \cite{KSST} 
possessing a new NL SUSY in curved spacetime. 
In particular, in the context of SGM 
it is inevitable to linearize NL SUSY 
and to obtain a linear SUSY invariant action which corresponds 
to the (NL SUSY invariant) SGM action including the higher derivative 
terms of the N-G fermion in order to derive the low energy physics of SGM. 
From these viewpoint, it is useful to recognize how a supermultiplet 
of linear SUSY is expressed in terms of a (NL SUSY invariant) 
higher derivative action of the N-G fermion 
in addition to the V-A action. 
Also, it is important to know the relation of the actions 
with the N-G fermion and the standard V-A action, 
i.e., the universality of actions with the N-G fermion 
as discussed already in Ref.\cite{HK}. 

In this letter, we focus on $N = 1$ SUSY for simplicity 
and investigate a relation between the scalar supermultiplet action 
of linear SUSY \cite{WZ1} and a NL SUSY model including 
apparently pathological higher derivative terms 
of the N-G fermion besides the V-A action. 
We show by heuristic arguments 
that SUSY invariant relations with higher derivative 
terms of the N-G fermion, which connect the linear 
and NL SUSY models, are constructed at leading orders 
starting from an ansatz with a higher (first-order) derivative 
term of the N-G fermion as given below in Eq.(\ref{ansatz}). 
We also briefly discuss a higher derivative 
action of the N-G fermion in the NL SUSY model. 
Our model includes higher derivative terms of the N-G fermion 
which apparently describe a (Weyl) ghost field. 
By using this relation, 
we also explicitly prove an equivalence between the standard 
NL SUSY V-A model and our NL SUSY model 
with the pathological higher derivatives of the N-G fermion. 
This is a different example from the arguments of \cite{HK} 
with respect to the universality of NL SUSY actions 
with the N-G fermion. 

Let us denote the component fields of a $N = 1$ scalar supermultiplet 
\cite{WZ1} as $(A, B, \lambda, F, G)$ 
\footnote{
In this letter Minkowski spacetime indices are denoted 
by $a, b, ... = 0, 1, 2, 3$, 
and we use the  Minkowski spacetime metric 
${1 \over 2}\{ \gamma^a, \gamma^b \} = \eta^{ab}= (+, -, -, -)$ 
and $\sigma^{ab} = {i \over 4}[\gamma^a, \gamma^b]$.}, 
in which $A$ and $B$ are two physical scalar fields, 
$\lambda$ is a Majorana spinor, 
and $F$ and $G$ means two auxiliary scalar fields. 
The linear SUSY transformations of these component fields 
generated by a constant (Majorana) spinor parametar $\zeta$ 
are written by 
\ba
\A \A 
\delta_Q A = \bar\zeta \lambda, \nonu
\A \A 
\delta_Q B = i \bar\zeta \gamma_5 \lambda, \nonu
\A \A 
\delta_Q \lambda = \{ (F + i \gamma_5 G) 
- i \!\!\not\!\partial (A + i \gamma_5 B) \} \zeta, \nonu
\A \A 
\delta_Q F = - i \bar\zeta \!\!\not\!\partial \lambda, \nonu
\A \A 
\delta_Q G = \bar\zeta \gamma_5 \!\!\not\!\partial \lambda. 
\label{LSUSY}
\ea
These transformations satisfy a closed off-shell commutator algebra, 
$[\delta_Q(\zeta_1), \delta_Q(\zeta_2)]$ $= \delta_P(v)$, 
where $\delta_P$ is a translation with a parameter 
$v^a = 2i \bar\zeta_1 \gamma^a \zeta_2$. 

On the other hand, for the $N = 1$ V-A model \cite{VA} 
we have a NL SUSY transformation law of a (Majorana) N-G fermion 
$\psi$ generated by $\zeta$, 
\be
\delta_Q \psi = {1 \over \kappa} \zeta 
- i \kappa (\bar\zeta \gamma^a \psi) \partial_a \psi, 
\label{NLSUSY}
\ee
where $\kappa$ is a constant whose dimension is $({\rm mass})^{-2}$, 
and Eq.(\ref{NLSUSY}) also satisfies the off-shell commutator algebra, 
$[\delta_Q(\zeta_1), \delta_Q(\zeta_2)] = \delta_P(v)$. 

As for the method of constructing SUSY invariant relations 
between the component fields of the $N = 1$ scalar supermultiplet 
and the N-G fermion field $\psi$, there is a heuristic method \cite{Ro} 
starting from an ansatz, $\lambda = \psi + {\cal O}(\kappa^2)$, 
and obtaining higher order terms such that the linear SUSY transformations 
(\ref{LSUSY}) are reproduced by using 
the NL SUSY transformation (\ref{NLSUSY}). 
In this letter, following this method 
and starting from the following ansatz 
with a higher (first-order) derivative term of $\psi$, 
\be
\lambda \ = \ \psi \ + \ i \kappa^{1 \over 2} \ \!\!\not\!\partial \psi 
+ \ {\cal O}(\kappa^2) \ + \ {\cal O}(\kappa^{5/2}), 
\label{ansatz}
\ee
we construct the SUSY invariant relations 
between the component fields of the $N = 1$ scalar supermultiplet 
and the N-G fermion field $\psi$ at leading orders. 

Indeed, after some calculations we obtain the relations between the fields 
of $N = 1$ scalar supermultiplet and the N-G fermion $\psi$ as 
\ba
A = \A \A 
{1 \over 2} \kappa \ \bar\psi \psi 
- {i \over 4} \kappa^3 \ 
\{ (\bar\psi \!\!\not\!\partial \psi) \bar\psi \psi 
- (\bar\psi \gamma_5 \!\!\not\!\partial \psi) \bar\psi 
\gamma_5 \psi \} + {\cal O}(\kappa^5) \nonu
\A \A 
+ i \kappa^{3 \over 2} \ \bar\psi \!\!\not\!\partial \psi 
+ \kappa^{7 \over 2} \ \left[ \ {1 \over 4} \partial_a 
\{ (\bar\psi \partial^a \psi) \bar\psi \psi 
- (\bar\psi \gamma_5 \partial^a \psi) \bar\psi \gamma_5 \psi \} \right. 
\nonu
\A \A 
- {i \over 2} \{ (\partial_a \bar\psi \sigma^{ab} \partial_b \psi) \bar\psi \psi 
- (\partial_a \bar\psi \gamma_5 \sigma^{ab} \partial_b \psi) 
\bar\psi \gamma_5 \psi \} 
\nonu
\A \A 
+ {i \over 2} \epsilon^{abcd} 
(\bar\psi \gamma_c \partial_a \psi) \bar\psi \gamma_5 \gamma_d \partial_b \psi 
\nonu
\A \A 
\left. 
+ {1 \over 2} \{ (\bar\psi \!\!\not\!\partial \psi) \bar\psi \!\!\not\!\partial \psi 
- (\bar\psi \gamma^a \partial_b \psi) \bar\psi \gamma^b \partial_a \psi \} 
\ \right] + {\cal O}(\kappa^{11 \over 2}), 
\label{const-A} \\[5mm]
B = \A \A 
{i \over 2} \kappa \ \bar\psi \gamma_5 \psi 
+ {1 \over 4} \kappa^3 \ 
\{ (\bar\psi \!\!\not\!\partial \psi) \bar\psi \gamma_5 \psi 
- (\bar\psi \gamma_5 \!\!\not\!\partial \psi) \bar\psi \psi \} 
+ {\cal O}(\kappa^5) 
\nonu
\A \A 
- \kappa^{3 \over 2} \ \bar\psi \gamma_5 \!\!\not\!\partial \psi 
+ i \kappa^{7 \over 2} \ \left[ \ {1 \over 4} \partial_a 
\{ (\bar\psi \partial^a \psi) \bar\psi \gamma_5 \psi 
- (\bar\psi \gamma_5 \partial^a \psi) \bar\psi \psi \} \right. 
\nonu
\A \A 
- {i \over 2} \{ (\partial_a \bar\psi \sigma^{ab} \partial_b \psi) 
\bar\psi \gamma_5 \psi 
- (\partial_a \bar\psi \gamma_5 \sigma^{ab} \partial_b \psi) \bar\psi \psi \} 
\nonu
\A \A 
+ {i \over 2} \epsilon^{abcd} 
(\bar\psi \gamma_5 \gamma_c \partial_a \psi) 
\bar\psi \gamma_5 \gamma_d \partial_b \psi 
\nonu
\A \A 
\left. 
+ {1 \over 2} \{ (\bar\psi \!\!\not\!\partial \psi) 
\bar\psi \gamma_5 \!\!\not\!\partial \psi 
- (\bar\psi \gamma^a \partial_b \psi) 
\bar\psi \gamma_5 \gamma^b \partial_a \psi \} 
\ \right] + {\cal O}(\kappa^{11 \over 2}), 
\ea
\ba
\lambda = \A \A \psi 
+ {i \over 2} \kappa^2 
\ \{ - (\bar\psi \!\!\not\!\partial \psi) \psi 
+ (\bar\psi \gamma_5 \!\!\not\!\partial \psi) \gamma_5 \psi 
- (\bar\psi \partial_a \psi) \gamma^a \psi 
- (\bar\psi \gamma_5 \partial_a \psi) \gamma_5 \gamma^a \psi \} \nonu
\A \A 
+ {\cal O}(\kappa^4) \nonu
\A \A 
+ i \kappa^{1 \over 2} \ \!\!\not\!\partial \psi \nonu
\A \A 
+ \kappa^{5 \over 2} \ \left[ \ {1 \over 2} 
\partial_a \{ (\bar\psi \partial^a \psi) \psi 
- (\bar\psi \gamma_5 \partial^a \psi) \gamma_5 \psi 
+ (\bar\psi \!\!\not\!\partial \psi) \gamma^a \psi 
+ (\bar\psi \gamma_5 \!\!\not\!\partial \psi) \gamma_5 \gamma^a \psi 
\} \right. \nonu
\A \A 
- i \{ (\partial_a \bar\psi \sigma^{ab} \partial_b \psi) \psi 
+ (\partial_a \bar\psi \gamma_5 \sigma^{ab} \partial_b \psi) \gamma_5 \psi \} 
\nonu
\A \A 
+ {i \over 2} \epsilon^{abcd} 
\{ (\bar\psi \gamma_5 \gamma_c \partial_a \psi) \gamma_d \partial_b \psi 
+ (\bar\psi \gamma_c \partial_a \psi) \gamma_5 \gamma_d \partial_b \psi \} 
\nonu
\A \A 
+ i \{ (\bar\psi \partial_a \psi) \sigma^{ab} \partial_b \psi 
- (\bar\psi \gamma_5 \partial_a \psi) \gamma_5 \sigma^{ab} \partial_b \psi \} 
\ \Bigg] + {\cal O}(\kappa^{9 \over 2}), 
\ea
\ba
F = \A \A 
\left( {1 \over \kappa} - i \kappa \bar\psi 
\!\!\not\!\partial \psi \right) 
+ {1 \over 2} \kappa^3 \ 
\left[ \ i \{ (\partial_a \bar\psi \sigma^{ab} \partial_b \psi) 
\bar\psi \psi 
- (\partial_a \bar\psi \gamma_5 \sigma^{ab} \partial_b \psi) 
\bar\psi \gamma_5 \psi \} \right. \nonu
\A \A 
- (\partial_a \bar\psi \gamma_5 \!\!\not\!\partial \psi) 
\bar\psi \gamma_5 \gamma^a \psi 
+ (\bar\psi \gamma^b \partial_a \psi) 
\bar\psi \gamma^a \partial_b \psi 
- (\bar\psi \gamma_5 \gamma^b \partial_a \psi) 
\bar\psi \gamma_5 \gamma^a \partial_b \psi \nonu
\A \A 
\left. - {1 \over 4} \{ (\bar\psi \psi) \Box (\bar\psi \psi) 
- (\bar\psi \gamma_5 \psi) \Box (\bar\psi \gamma_5 \psi) 
\} \ \right] + {\cal O}(\kappa^5) \nonu
\A \A 
+ \kappa^{3 \over 2} \ \partial_a 
(\bar\psi \gamma^a \!\!\not\!\partial \psi) 
- i \kappa^{7 \over 2} 
\ \left[ \ {1 \over 4} \{ \Box (\bar\psi \!\!\not\!\partial \psi) \bar\psi \psi 
- \Box (\bar\psi \gamma_5 \!\!\not\!\partial \psi) \bar\psi \gamma_5 \psi \} \right. 
\nonu
\A \A 
+ \left( {1 \over 4} \partial_a \bar\psi \gamma_5 \gamma^a \gamma_b \Box \psi 
+ {1 \over 2} \partial_a \partial_c \bar\psi \gamma_5 \gamma^c \gamma_b \partial^a \psi 
- 2i \partial_a \partial_b \bar\psi \gamma_5 \sigma^{ac} \partial_c \psi 
\right) \bar\psi \gamma_5 \gamma^b \psi 
\nonu
\A \A 
+ (\partial_a \bar\psi \!\!\not\!\partial \psi) \bar\psi \partial^a \psi 
- (\partial_a \bar\psi \gamma_5 \!\!\not\!\partial \psi) 
\bar\psi \gamma_5 \partial^a \psi 
- (\partial_a \bar\psi \partial_b \psi) \bar\psi \gamma^a \partial^b \psi 
\nonu
\A \A 
+ (\partial_a \bar\psi \gamma_5 \partial_b \psi) 
\bar\psi \gamma_5 \gamma^a \partial^b \psi 
- i \{ (\partial_a \bar\psi \sigma^{ab} \partial_b \psi) 
\bar\psi \!\!\not\!\partial \psi 
+ (\partial_a \bar\psi \gamma_5 \sigma^{ab} \partial_b \psi) 
\bar\psi \gamma_5 \!\!\not\!\partial \psi \} 
\nonu
\A \A 
+ {i \over 2} \epsilon^{abcd} 
\{ (\partial_e \bar\psi \gamma_5 \gamma_c \partial_a \psi) 
\bar\psi \gamma^e \gamma_d \partial_b \psi 
- (\partial_e \bar\psi \gamma_c \partial_a \psi) 
\bar\psi \gamma_5 \gamma^e \gamma_d \partial_b \psi \} 
\nonu
\A \A 
+ 4i (\partial_a \bar\psi \sigma^{bc} \partial_c \psi) 
\bar\psi \gamma^a \partial_b \psi \ \Bigg] + {\cal O}(\kappa^{11 \over 2}), 
\label{const-F}
\ea
\ba
G = \A \A 
\kappa \ \bar\psi \gamma_5 \!\!\not\!\partial \psi 
- {i \over 2} \kappa^3 \ 
\left[ \ i \{ (\partial_a \bar\psi \gamma_5 \sigma^{ab} \partial_b \psi) 
\bar\psi \psi 
- (\partial_a \bar\psi \sigma^{ab} \partial_b \psi) 
\bar\psi \gamma_5 \psi \} \right. 
\nonu
\A \A 
\left. + (\partial_a \bar\psi \!\!\not\!\partial \psi) 
\bar\psi \gamma_5 \gamma^a \psi 
+ {1 \over 4} \{ (\bar\psi \gamma_5 \psi) \Box (\bar\psi \psi) 
- (\bar\psi \psi) \Box (\bar\psi \gamma_5 \psi) 
\} \ \right] + {\cal O}(\kappa^5) 
\nonu
\A \A 
+ i \kappa^{3 \over 2} \ \partial_a 
(\bar\psi \gamma_5 \gamma^a \!\!\not\!\partial \psi) 
+ \kappa^{7 \over 2} 
\ \left[ \ {1 \over 4} \{ \Box (\bar\psi \!\!\not\!\partial \psi) \bar\psi \gamma_5 \psi 
- \Box (\bar\psi \gamma_5 \!\!\not\!\partial \psi) \bar\psi \psi \} \right. 
\nonu
\A \A 
+ \left( {1 \over 4} \partial_a \bar\psi \gamma^a \gamma_b \Box \psi 
+ {1 \over 2} \partial_a \partial_c \bar\psi \gamma^c \gamma_b \partial^a \psi 
- 2i \partial_a \partial_b \bar\psi \sigma^{ac} \partial_c \psi 
\right) \bar\psi \gamma_5 \gamma^b \psi 
\nonu
\A \A 
+ (\partial_a \bar\psi \!\!\not\!\partial \psi) \bar\psi \gamma_5 \partial^a \psi 
- (\partial_a \bar\psi \gamma_5 \!\!\not\!\partial \psi) \bar\psi \partial^a \psi 
+ (\partial_a \bar\psi \partial_b \psi) \bar\psi \gamma_5 \gamma^a \partial^b \psi 
\nonu
\A \A 
- (\partial_a \bar\psi \gamma_5 \partial_b \psi) \bar\psi \gamma^a \partial^b \psi 
- i \{ (\partial_a \bar\psi \sigma^{ab} \partial_b \psi) 
\bar\psi \gamma_5 \!\!\not\!\partial \psi 
+ (\partial_a \bar\psi \gamma_5 \sigma^{ab} \partial_b \psi) 
\bar\psi \!\!\not\!\partial \psi \} 
\nonu
\A \A 
+ {i \over 2} \epsilon^{abcd} 
\{ (\partial_e \bar\psi \gamma_5 \gamma_c \partial_a \psi) 
\bar\psi \gamma_5 \gamma^e \gamma_d \partial_b \psi 
- (\partial_e \bar\psi \gamma_c \partial_a \psi) 
\bar\psi \gamma^e \gamma_d \partial_b \psi \} 
\nonu
\A \A 
+ 4i (\partial_a \bar\psi \gamma_5 \sigma^{bc} \partial_c \psi) 
\bar\psi \gamma^a \partial_b \psi \ \Bigg] + {\cal O}(\kappa^{11 \over 2}). 
\label{const-G}
\ea
It is straightforward but lengthy to prove that the linear SUSY transformations 
(\ref{LSUSY}) are reproduced by using the NL SUSY transformation (\ref{NLSUSY}). 
The SUSY invariant relations at ${\cal O}(\kappa^{2m})$\ ($m = 0, 1, 2, \cdots$) 
or ${\cal O}(\kappa^{2m+1})$ in Eqs. from (\ref{const-A}) to (\ref{const-G}) 
are those obtained in \cite{IK}-\cite{UZ}. 
And also the relation (\ref{const-F}) at ${\cal O}(\kappa^{2m+1})$ 
for the auxiliary field $F$ 
has the form which is proportional to a determinant 
$\vert w \vert = {\rm det}(w^a{}_b)$ in $N = 1$ V-A model \cite{VA} 
with $w^a{}_b$ being defined by 
\be
w^a{}_b = \delta^a{}_b + t^a{}_b, 
\ \ \ t^a{}_b = - i \kappa^2 \bar\psi \gamma^a \partial_b \psi, 
\ee
plus total derivative terms \cite{Ro}; 
namely, Eq.(\ref{const-F}) becomes 
$F = (1/\kappa) \vert w \vert + [\ {\rm tot.\ der.}\ ]$ 
which shows that $1/\kappa$ corresponds to the vacuum 
expectation value of the auxiliary field $F$. 
The SUSY invariant relations at 
${\cal O}(\kappa^{2m+{1 \over 2}})$ or ${\cal O}(\kappa^{2m+{3 \over 2}})$ 
are the new higher derivative terms of the N-G fermion. 

The derivation of the above SUSY invariant relations 
from (\ref{const-A}) to (\ref{const-G}) does not depend on the form 
of the action for the linear and NL SUSY models. 
We now consider a free action which is invariant under Eq.(\ref{LSUSY}) 
\be
S_{\rm lin} = 
\int d^4 x 
\left[ {1 \over 2} (\partial_a A)^2 
+ {1 \over 2} (\partial_a B)^2 
+ {i \over 2} \bar\lambda \!\!\not\!\partial \lambda 
+ {1 \over 2} (F^2 + G^2) 
- {1 \over \kappa} F \right]. 
\label{Lact}
\ee
The last term proportional to $\kappa^{-1}$ 
is an analog of the Fayet-Iliopoulos $D$ term in the $N = 1$ 
gauge supermultiplet \cite{FI}. 
The field equations for the auxiliary field $F$ 
is $F = 1/\kappa$ indicating a spontaneous SUSY breaking. 
As already shown in \cite{IK}-\cite{UZ}, 
substituting the terms at ${\cal O}(\kappa^{2m})$ and ${\cal O}(\kappa^{2m+1})$ 
of Eqs. from (\ref{const-A}) to (\ref{const-G}) 
into the linear action $S_{\rm lin}$ of Eq.(\ref{Lact}) 
gives the V-A action $S_{\rm VA}$, 
\ba
S_{\rm VA} = \A \A - {1 \over {2 \kappa^2}} 
\int d^4 x \ \vert w \vert \nonu
= \A \A 
- {1 \over {2 \kappa^2}} \int d^4 x 
\left[ 1 + t{^a}_a 
+ {1 \over 2}(t{^a}_a t{^b}_b - t{^a}_b t{^b}_a) \right. \nonu
\A \A 
\left. - {1 \over 6} \epsilon_{abcd} \epsilon^{efgd} t{^a}_e t{^b}_f t{^c}_g 
- {1 \over 4!} \epsilon_{abcd} \epsilon^{efgh} t{^a}_e t{^b}_f t{^c}_g t{^d}_h 
\right], 
\label{VAact}
\ea
which is invariant under the NL SUSY transformation (\ref{NLSUSY}). 

When we substitute the terms at both (${\cal O}(\kappa^{2m})$, ${\cal O}(\kappa^{2m+1})$) 
and (${\cal O}(\kappa^{2m+{1 \over 2}})$, ${\cal O}(\kappa^{2m+{3 \over 2}})$) 
of Eqs. from (\ref{const-A}) to (\ref{const-G}) 
into the linear action $S_{\rm lin}$ of Eq.(\ref{Lact}), 
$S_{\rm lin}$ leads a higher derivative action $S_{\rm higher\ der}$ 
with respect to $\psi$ in addition to the V-A action $S_{\rm VA}$ of Eq.(\ref{VAact}); 
namely, $S_{\rm lin} = S_{\rm VA} + S_{\rm higher\ der}$, 
and $S_{\rm higher\ der}$ at ${\cal O}(\psi^2)$ (i.e. up to ${\cal O}(\kappa)$) 
is given for example as 
\be
S_{\rm higher\ der} [{\cal O}(\psi^2)] = \int d^4 x 
\left[ \kappa^{1/2} \partial_a \bar\psi \partial^a \psi 
+ {i \over 2} \kappa \partial_a \bar\psi \gamma^a \Box \psi \right] 
\label{hdact}
\ee
except for total derivative terms. 
Such terms as in Eq.(\ref{hdact}) at ${\cal O}(\psi^2)$ 
are discussed in the context of a higher derivative fermionic field theory 
(for example, see \cite{Vi}), in which a (Weyl) ghost field is included. 
Higher order terms of $\psi$ in $S_{\rm higher\ der}$, 
e.g., terms at ${\cal O}(\psi^4)$ can be obtained 
from the results in Eqs. from (\ref{const-A}) to (\ref{const-G}). 

In the above consideration, the apparently pathological 
higher derivative terms (\ref{hdact}) 
appear in the NL SUSY model which is described by 
$S_{\rm VA} + S_{\rm higher\ der}$. 
However, we also find that 
the above NL SUSY model, $S_{\rm VA} + S_{\rm higher\ der}$, 
is equivalent to the standard NL SUSY V-A model described only 
by $S_{\rm VA}$ of Eq.(\ref{VAact}). 
In order to show this, 
here we explicitly construct the {\it NL SUSY invariant} 
relation which connects the two NL SUSY actions, 
$S_{\rm VA} + S_{\rm higher\ der}$ and $S_{\rm VA}$. 

Indeed, let us denote the N-G fermion field of 
the standard V-A model as $\psi'$, 
i.e., $S_{\rm VA} = S_{\rm VA}[\psi']$ which is invariant under 
the NL SUSY transfomation generated by $\zeta$, 
\be
\delta_Q \psi' = {1 \over \kappa} \zeta 
- i \kappa (\bar\zeta \gamma^a \psi') \partial_a \psi'. 
\label{NLSUSY0}
\ee
The form of NL SUSY transformation law (\ref{NLSUSY0}) 
is the same as Eq.(\ref{NLSUSY}). 
We also consider the relation between $S_{\rm VA}[\psi']$ 
and $(S_{\rm VA} + S_{\rm higher\ der})[\psi]$, 
starting from the following ansatz, 
\be
\psi' \ = \ \psi \ + \ i \kappa^{1 \over 2} \ \!\!\not\!\partial \psi 
+ \ {\cal O}(\kappa^{5/2}), 
\label{ansatz0}
\ee
in order to derive Eq.(\ref{hdact}) in the NL SUSY model, 
$(S_{\rm VA} + S_{\rm higher\ der})[\psi]$. 
Then we easily obtain next higher order terms in Eq.(\ref{ansatz0}) 
such that the NL SUSY transformation (\ref{NLSUSY0}) 
is reproduced by Eq.(\ref{NLSUSY}); namely, we have 
\be
\psi' = \psi + i \kappa^{1 \over 2} \!\!\not\!\partial \psi 
+ \kappa^{5 \over 2} (\bar\psi \gamma^a \!\!\not\!\partial \psi \partial_a \psi 
- \bar\psi \gamma^a \partial_b \psi \gamma^b \partial_a \psi) 
+ i \kappa^3 \bar\psi \gamma^a \!\!\not\!\partial \psi 
\partial_a \!\!\not\!\partial \psi + {\cal O}(\kappa^{9/2}). 
\label{const-p'}
\ee
Here we can further continue to obtain higher order terms 
in the SUSY invariant relation (\ref{const-p'}). 
When we substitute Eq.(\ref{const-p'}) into the standard V-A action 
$S_{\rm VA}[\psi']$, the $S_{\rm VA}[\psi']$ exactly leads 
to Eq.(\ref{hdact}) (up to the total derivertive terms 
which have been omitted from Eq.(\ref{hdact})) 
in addition to the $S_{\rm VA}[\psi]$. 
We also expect that the $S_{\rm VA}[\psi']$ leads to 
higher order terms of $\psi$ in $S_{\rm higher\ der}[\psi]$, 
e.g., terms at ${\cal O}(\psi^4)$. 

We summarize the results as follows. 
Adopting the ansatz (\ref{ansatz}) 
with the higher (first-order) derivative term of the N-G fermion, 
we have investigated for $N = 1$ SUSY the relation 
between the scalar supermultiplet of linear SUSY 
and the NL SUSY model including apparently pathological 
higher derivative terms of the N-G fermion besides the V-A action. 
We have explicitly shown that 
the component fields of the $N = 1$ scalar supermultiplet 
are consistently expanded in terms of the N-G fermion 
in the NL SUSY invariant way as Eqs. from (\ref{const-A}) to (\ref{const-G}). 
The (NL SUSY invariant) higher derivative action of the N-G fermion, 
which apparently includes a (Weyl) ghost field, 
has been discussed at the leading order in Eq.(\ref{hdact}). 
By using this relation 
and by constructing the NL SUSY invariant relation (\ref{const-p'}), 
we have also explicitly proved the equivalence between the standard 
NL SUSY V-A model and our NL SUSY model 
with the pathological higher derivatives 
as an example with respect to the universality of NL SUSY actions 
with the N-G fermion. 
In other words, the NL SUSY invariant condition may give 
more general field redefinitions, 
which reproduce the standard V-A model.

%
%

\vspace{10mm}

\noindent
{\Large{\bf Acknowledgements}} \\[3mm]
We are grateful to Professor Sergei V. Ketov 
for the interest in our work 
and for valuable suggestions and comments.

\newpage

%
\newcommand{\NP}[1]{{\it Nucl.\ Phys.\ }{\bf #1}}
\newcommand{\PL}[1]{{\it Phys.\ Lett.\ }{\bf #1}}
\newcommand{\CMP}[1]{{\it Commun.\ Math.\ Phys.\ }{\bf #1}}
\newcommand{\MPL}[1]{{\it Mod.\ Phys.\ Lett.\ }{\bf #1}}
\newcommand{\IJMP}[1]{{\it Int.\ J. Mod.\ Phys.\ }{\bf #1}}
\newcommand{\PR}[1]{{\it Phys.\ Rev.\ }{\bf #1}}
\newcommand{\PRL}[1]{{\it Phys.\ Rev.\ Lett.\ }{\bf #1}}
\newcommand{\PTP}[1]{{\it Prog.\ Theor.\ Phys.\ }{\bf #1}}
\newcommand{\PTPS}[1]{{\it Prog.\ Theor.\ Phys.\ Suppl.\ }{\bf #1}}
\newcommand{\AP}[1]{{\it Ann.\ Phys.\ }{\bf #1}}

\end{document}